\documentclass[12pt]{article}
\pdfoutput=1
\usepackage{graphicx,lutypaper}
\numberwithin{equation}{section} 

\begin{document}


\begin{titlepage}

\title{The Higgs Trilinear Coupling\\
and the Scale of New Physics}

\author{Spencer Chang}

\address{Department of Physics and Institute of Theoretical Science\\ 
University of Oregon, Eugene, Oregon 97403\\
{\normalfont and}\\
Department of Physics \\  
National Taiwan University, Taipei, Taiwan 10617, ROC}

\author{Markus A. Luty}

\address{Center for Quantum Mathematics and Physics (QMAP)\\
University of California, Davis, California 95616}

\begin{abstract}
We consider modifications of the Higgs potential due to new physics 
at high energy scales.
These upset delicate cancellations predicted by the Standard Model
for processes involving Higgs bosons and longitudinal gauge bosons, and lead to
a breakdown of the theory at high energies.
We focus on modifications of the Higgs trilinear coupling and use the
violation of tree-level unitarity as an estimate of the
scale where the theory breaks down.
We obtain a completely model-independent bound of $\lsim 13\TeV$ for an
order-1 modification of the trilinear.
We argue that this bound can be saturated only in fine-tuned models, and
the scale of new physics is likely to be much lower.
The most stringent bounds are obtained from 
amplitudes involving multiparticle states that are not conventional scattering states.
Our results show that a future determination of the Higgs cubic coupling
can point to a well-defined scale of new physics that can be targeted and explored 
at future colliders.
\end{abstract}

\end{titlepage}

\noindent
\section{Introduction}
\begin{figure*}[h]
\begin{center}
\begin{minipage}{5.75in}
\begin{center}
\includegraphics{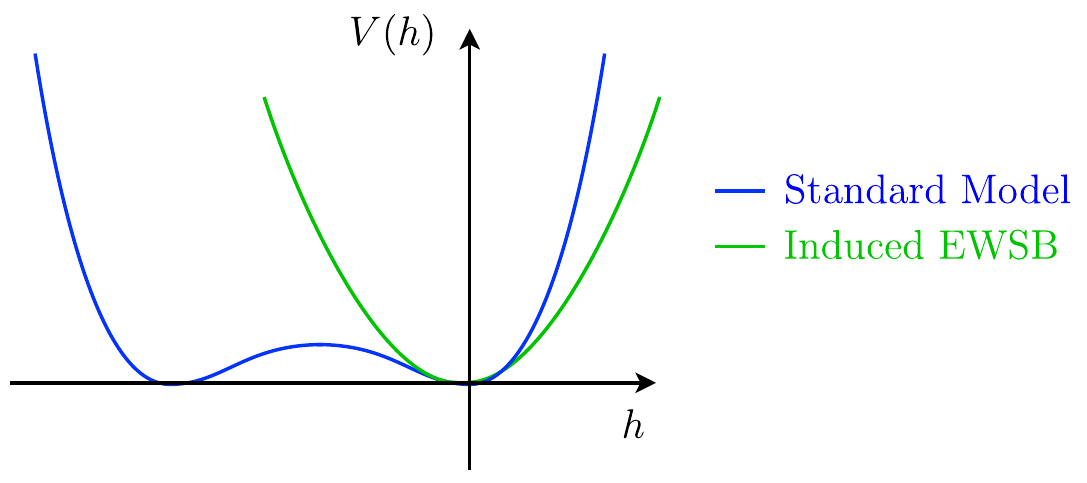}
\small\caption{Higgs potentials in the Standard Model 
and in induced electroweak symmetry breaking.
\label{Fig:Potentials}}
\end{center}
\end{minipage}
\end{center}
\end{figure*}

Many of the couplings of the 125~GeV Higgs boson to gauge bosons and fermions have 
been measured at the $10\%$ level and agree with the predictions of the
Standard Model \cite{ATLAS-CONF-2018-031,Sirunyan:2018koj}.
On the other hand, the Higgs potential is very weakly constrained
experimentally.
If we define $h$ to be the Higgs field measured relative to its vacuum
expectation value, the $h^2$ term gives the Higgs mass, but the higher
order terms in the potential are 
very weakly constrained.
For example, we currently do not know whether the Higgs potential is a
double well as predicted by the Standard Model, or a shifted single-well
as in models of induced electroweak symmetry breaking \cite{Azatov:2011ht,Azatov:2011ps,Galloway:2013dma,Chang:2014ida}
(see Fig.~\ref{Fig:Potentials}).  Another well motivated theory is the Standard Model with a large, modified Higgs trilinear, giving the strong first order electroweak phase transition as needed for electroweak baryogenesis \cite{Grojean:2004xa}.
Such models can be clearly distinguished by the coefficient of the $h^3$ term in the Higgs
potential, which can be probed in di-Higgs production.
This measurement is difficult due to low rates and large backgrounds.  
The present limits from the LHC 
constrain the Higgs trilinear to lie in the range $-5$ to $+12$ times 
the Standard Model value \cite{Aad:2019uzh,ATLAS-CONF-2019-049,Sirunyan:2018two}. 
Current studies for the high-luminosity LHC indicate that Higgs pair production can only probe the trilinear coupling at best at the
$O(1)$ level  \cite{ATL-PHYS-PUB-2017-001,CMS:2015nat, Cepeda:2019klc}.
Future high energy lepton or hadron colliders are required for a more accurate determination with potential sensitivity at the 10\% level
\cite{Duerig:2016dvi,Abramowicz:2016zbo,Goncalves:2018yva}.  
In comparison, even at future colliders,  triple Higgs production is not sensitive to the Standard Model prediction, but can be sensitive to large enough modifications \cite{Papaefstathiou:2015paa,Chen:2015gva, Belyaev:2018fky}.    
For a recent review on collider Higgs probes, see~\cite{Dawson:2018dcd}.  
Indirect constraints on the Higgs self interactions have also been studied for precision 
electroweak observables and loop level corrections to Higgs cross sections 
({\it e.g.}\cite{Degrassi:2017ucl, Kribs:2017znd,McCullough:2013rea, Gorbahn:2016uoy, 
Degrassi:2016wml, Liu:2018peg, Bizon:2018syu, Borowka:2018pxx}), 
which also have sensitivity but are more model dependent.

An important motivation for this difficult measurement 
is that a deviation from the Standard Model prediction for
the Higgs cubic coupling is a sign of new fundamental particles and/or interactions
beyond those described by the Standard Model.
By itself, the deviation in one coupling will not give very much information
about what kind of new physics is responsible.
However, one model-independent conclusion that can be drawn from such a result
is that the mass scale associated with this new physics  
cannot be arbitrarily large.
This is because the Standard Model is the unique perturbatively
UV complete theory containing only the experimentally observed
elementary particles and interactions.
If the new physics that gives rise to the deviation occurs at a mass scale $M$
that is much larger than $m_h$, then physics below the scale $M$ can be
described by an effective theory with the same degrees of freedom as the 
Standard Model.
This theory will not be UV complete, and will break down at some UV scale,
which in turn gives an upper bound on the scale $M$.
This bound can be determined entirely from the effective theory, which 
consists of the Standard Model plus additional local terms that account for
the observed deviation from the Standard Model.

In this paper, we give a model-independent estimate of the scale of new physics 
associated with a deviation in the measured value of the Higgs cubic interaction.  
The point is that the UV incompleteness of the effective theory that describes 
the deviation shows up in the violation of tree-level unitarity at high energies.
Unitarity is restored order by order in perturbation theory, since the Hamiltonian is Hermitian,
but the violation of unitarity at tree-level means that the loop corrections that restore unitarity are
order-1 corrections, {\it i.e.}~the loop expansion is breaking down and the theory is becoming strongly coupled.  We interpret this strong coupling as a sign that new physics is required at or below this
scale, since there is no unique way to extrapolate the theory to higher energy scales.
This is the 
standard argument that leads to the use of tree-level unitarity
as a diagnostic for the scale of new physics.
Note that the scale of strong coupling 
defined in this way
is ambiguous up to $O(1)$ factors;
for example loop corrections near the strong coupling scale will give $O(1)$ corrections
to tree-level couplings.
Nonetheless, tree-level unitarity gives a useful $O(1)$ estimate of the
scale where we expect new physics 
to appear.

This is a variation on a classic success story in particle physics.
Long before the discovery of the Higgs boson,
Lee, Quigg, and Thacker \cite{Lee:1977eg, PhysRevLett.38.883}
observed that in a theory without a Higgs sector,
the scattering of longitudinally polarized $W$ and $Z$ bosons violates 
tree-level unitarity in the UV, and used this to give a model-independent upper 
bound on the scale  of the Higgs sector.
This work has been refined and extended in many ways, see for example \cite{Dicus:1992vj, Chanowitz:1985hj}.  
These arguments were 
one of the most important motivations for the energy scale of the LHC,
which was in fact successful in discovering the Higgs boson.
The present experimental situation is in a sense opposite, in that the minimal 
theory with the observed particle content can be consistently extrapolated 
to exponentially large energy scales.
But this is only the case if all of the couplings 
in the theory have the precise values predicted
by the Standard Model.
Any observed deviation from the predictions of the Standard Model will therefore point
to a scale of new physics, just as the existence of massive $W$ and $Z$ bosons 
pointed to the scale of the Higgs sector.

There are several novel features in our analysis.
First, we do not assume that the leading deviations from the Standard Model
arise from the leading terms in an effective field theory framework
such as SMEFT \cite{Buchmuller:1985jz,Grzadkowski:2010es} 
or HEFT \cite{Grinstein:2007iv} (as reviewed in \cite{deFlorian:2016spz,Brivio:2017vri}).
These are different parameterizations of the most general physics beyond
the Standard Model, assuming no new light particles.
They differ only in the power counting rule that determines the relative
importance of various contributions.
In other words, a given deviation from the SM may appear at different orders in
the SMEFT and HEFT expansions, so the truncation of the expansion to a finite number
of terms can be different.
There is no universally correct expansion;
various types of new physics have effective theories with different
low-energy expansions
(see for example \cite{Giudice:2007fh}).
In this work we will be maximally conservative, and simply assume that a measured
deviation in the Higgs cubic coupling could come with any combination of
the infinitely many interaction terms that we can add to the effective theory.
That is, we allow arbitrary cancellations or conspiracies among these infinitely
many terms.
We simply maximize the scale of new physics over all possible ways of
accounting for the deviation, so our result is 
completely model-independent.

There are several other innovations in our analysis of a more
technical nature that improve upon earlier studies of unitarity violation
in many-particle amplitudes \cite{Belyaev:2012bm}.
First, we give a simple method of identifying the leading unitarity-violating
processes implied by a given local modification of the Standard Model
Lagrangian by using the equivalence theorem.
Another novelty is that the strongest model independent unitarity bounds on 
a Higgs trilinear arise from
3-to-3 processes, and we show that in general the optimal unitarity bounds 
result from using states other than conventional scattering states.

We comment briefly on some related work.
\Ref{Belyaev:2012bm} also studied high-energy unitarity violation in 
multiparticle processes arising from deviations in the Higgs potential,
focusing mainly on the possibility of experimentally observing large 
$2 \to n$ processes.
Some more recent work along these lines is in \Ref{Henning:2018kys}.
\Ref{DiLuzio:2017tfn} looked at unitarity violation due to the Higgs trilinear in Higgs-Higgs scattering, which for a 
sufficiently large
deviation, violates unitarity at low energies rather than at high energies.
Constraints from perturbative unitarity for $2\to 2$ processes in the context of SMEFT were studied in
\Refs{Corbett:2014ora,Corbett:2017qgl}, but not for processes induced by 
new contributions to the Higgs potential.
Recent work has also raised the possibility of large non-perturbative effects
within the Standard Model related to processes with large multiplicity produced at threshold
(for example \Ref{Jaeckel:2014lya}).
The size of these effects
is controversial \cite{Khoze:2017tjt,Monin:2018cbi}.
In any case, our
work involves processes with low multiplicity in the highly boosted regime which are under perturbative
control in the Standard Model, but that violate unitarity due to deviations
from the Standard Model.

This paper is organized as follows. 
In \S\ref{sec:PotentialUnitarity} we analyze the constraints from unitarity 
violation from arbitrary modifications of the Higgs potential.  
We show that a generic modification of the Higgs potential leads to unitarity
violation near the TeV scale.
We also show that certain unitarity-violating processes are determined completely
by the deviation of the Higgs trilinear from the Standard Model prediction,
leading to a much more conservative but completely model-independent bound
near $13$~TeV for an order-1 deviation in the Higgs trilinear.
In \S\ref{sec:TrilinearModels}
we consider possible models of new physics at high energy scales.
We argue that models that saturate the conservative bound require fine-tuning of
not only the Higgs mass, but also the Higgs quartic coupling.
Our conclusions are in \S\ref{sec:conclusions},
and an appendix describes a general analysis of unitarity violation 
from scalar potential terms.

\section{Higgs Potential Modifications and Unitarity Violation}
\label{sec:PotentialUnitarity}
We assume that physics below some UV scale $\La$ can be described by an
effective quantum field theory with the particle content of the SM.
Any deviations from the SM therefore arise by integrating out
states above the scale $\La$, and will result in additional local couplings
of the SM fields.
As discussed in the introduction, we do not want to make any assumption about
the relative importance of the infinitely many possible terms in this
Lagrangian.
We simply assume that the effective Lagrangian can be written as
\[
\scr{L}_\text{eff} = \scr{L}_\text{SM} + \de \scr{L},
\]
where $\de\scr{L}$ is small only in the sense that the effective Lagrangian
agrees with experiment.
It is not {\it a priori} obvious that a framework with this level of generality has any predictive
power.
We begin by considering the case of modifications of the Higgs potential.
To perform a complete analysis, we have to include all allowed Higgs interactions,
including interactions with Standard Model gauge bosons
and fermions.
For example, di-Higgs production {\it via} gluon-gluon fusion depends on the
Higgs couplings to the top quark and to gluons, as well as the trilinear
Higgs coupling (see {\it e.g.}~\cite{Azatov:2015oxa}).
However, the couplings to gluons and top quarks have stronger experimental
constraints than the Higgs trilinear.
Derivative couplings must also be considered, but they typically give rise
to lower unitarity violating scales, as we will discuss at the end of this section.

To write the most general deviation in the Higgs potential, we can write the
Lagrangian in unitary gauge where the eaten Nambu-Goldstone bosons are
set to zero.
In this gauge, the Standard Model Higgs doublet is given by
\[
H = \frac{1}{\sqrt{2}} \mat{0 \\ v + h},
\]
where $h$ is the physical Higgs field.
We assume that the minimum of the Higgs potential is at $h = 0$,
{\it i.e.}~$v$ is the minimum of the full Higgs potential,
including any deviations from the SM.
The %
scalar
potential is therefore a function of $h$ 
alone:
\[
V_\text{eff} = V_\text{SM}(h) + \de V(h),
\]
where
\begin{subequations}
\eql{Vh}
\[
V_{\text{SM}}(h) &= \sfrac 12 m_h^2 h^2 
+  \frac{m_h^2}{2v} h^3 + \frac{m_h^2}{8v^2} h^4,
\\
\eql{deVh}
\de V(h) &= \sum_{n \, = \, 3}^\infty \frac{\de \la_n}{n!} h^n.
\]
\end{subequations}
Our focus on general Higgs boson interactions is reminiscent of the Higgs Effective 
Field Theory (HEFT) approach \cite{Grinstein:2007iv}.
Our assumption that $v$ is the true Higgs VEV implies that there are no $O(h)$
terms in $\de V$.
We do not include $O(h^2)$ terms in $\de V$ because these can be absorbed
into a redefinition of $m_h$, which is well measured.  
It is only the cubic and higher terms that represent a true deviation from
the SM, as opposed to a change in the value of SM parameters.
We have $v = 246$~GeV and $m_h = 125\GeV$ to high accuracy,
so the Higgs cubic and quartic couplings are accurately predicted
in the SM.

We comment briefly on the role of loop corrections to the Higgs potential.
Loops involving heavy particles beyond the Standard Model can be expanded in powers of
$h$ about the VEV to give local terms of the form given above.
This leaves only loops involving Standard Model fields.
These loop corrections are perturbatively small until we get to the scale
of tree-level unitarity violation.
At that scale, they are expected to be $O(1)$ corrections, so including them
would only change the unitarity violating scale by $O(1)$.

Gauge invariance is not manifest in \Eqs{Vh}, but the potential is
the same as a general gauge invariant potential
written in terms of the Higgs doublet $H$ when both potentials are expanded around the Higgs VEV. 
We can simply write the potential as a sum of gauge invariant
terms of the form $(H^\dagger H)^n$.
It is convenient to write this in terms of the variable
\[
Y &= H^\dagger H - \sfrac 12 v^2.
\]
This has vanishing VEV and is linear in $h$, 
so minimization of the potential is simply the
statement that the potential starts with a
positive $Y^2$ term.
The Higgs potential can then be written as
\begin{subequations}
\eql{VH}
\[
V_\text{SM} &= m_H^2 H^\dagger H + \la (H^\dagger H)^2
= \la Y^2 + \text{constant},
\\
\eql{deVH}
\de V &= \sum_{n \, = \, 3}^\infty \frac{c_n}{n!} Y^n,
\]
\end{subequations}
where $m_h^2 = 2\la v^2$.
The relation between the couplings $c_n$ and $\la_n$ is easily worked out:
\[
\de\la_3 = c_3 v^3,
\quad
\de \la_4 = c_4 v^4 + 6 c_3 v^2,
\quad\ldots. \eql{smeftmods}
\]
\Eqs{Vh} and \eq{VH}
are two completely equivalent parameterizations of the Higgs potential.
We are free to use the parameterization that is more useful for our
purposes.

The parameterization \Eq{VH} is the one most commonly used 
in discussions of new high-scale physics,
and it is worth recalling the reason for this.
If we make the assumption that the new physics is associated with
a large mass scale $M$, and that this physics decouples from
electroweak symmetry breaking in the limit where $M$ is large,
then we expect that the importance of the couplings in the expansion \Eq{VH} 
will be ordered by dimensional analysis:
\[
c_n \sim \frac{1}{M^{2(n-2)}}.
\eql{smeftpowercounting}
\]
More generally, we can write the most general term in the effective
Lagrangian as a sum of local gauge invariant operators, and assign
them a power of $M$ by dimensional analysis.
Assuming that all coefficients in units of $M$ are the same order,
we obtain a predictive truncation of the effective theory, 
the so-called ``Standard Model Effective Field Theory'' 
(SMEFT) \cite{Brivio:2017vri}.
Although this power counting is expected to hold in a large class of
models, it is not completely general.
For example, operators containing derivatives may be suppressed by 
a parametrically different scale \cite{Liu:2016idz}.
Alternatively,
the new physics may include heavy particles whose mass comes from electroweak
symmetry breaking, which do not decouple at large mass.
(Although the precision electroweak $S$ parameter is a nontrivial constraint
on such a scenario, it is not cleanly ruled out unless there are many
heavy non-decoupling states.)
Finally, we may want to allow some of the coefficients to be anomalously
small, perhaps because of weak couplings, accidental cancellations,
or approximate symmetries.

Our goal is to bound the scale of tree-level unitarity violation
associated with deviations from the SM in a completely model-independent way, 
focusing on the Higgs trilinear.%
\footnote{Previous analyses of the unitarity violation of the Higgs trilinear 
\cite{DiLuzio:2017tfn}, found that a large trilinear (about seven times the Standard 
Model value) leads to unitarity violation for the process $hh\to hh$ near threshold,
but with good behavior at high energies.  
Any new physics that can unitarize this process should therefore be at low
energies and within reach of the LHC.}
In doing so, we will not make any assumption about the relative size of 
the coefficients in the effective theory.
It
is most convenient 
for our purposes
to use the parameterization in \Eqs{Vh}, since the new terms in the
Lagrangian are in one-to-one correspondence with new effective couplings
at the weak scale.
Not surprisingly, we will find that
tree-level unitarity violation is dominated by 
amplitudes involving Higgs bosons and longitudinally polarized $W$ and $Z$ bosons
($W_L$ and $Z_L$). 
The equivalence theorem tells us that the scattering amplitudes of
$W_L$ and $Z_L$ at high energies are the same as the scattering amplitudes of 
the corresponding unphysical Nambu-Goldstone bosons in a general gauge.
We can determine the dependence on the unphysical Nambu-Goldstone bosons
directly from the potential \Eq{deVh} by using the 
gauge invariant operator
\[
X &= \sqrt{2 H^\dagger H} - v
\nn
\eql{theX}
&= h +  \frac{\vec{G}^2}{2(v+h)} - \frac{\vec{G}^4}{8(v+h)^3}
+ O\left( \frac{\vec{G}^6}{(v+h)^5} \right),
\]
where $\vec{G} = (G_1, G_2, G_3)$ are the unphysical Nambu-Goldstone
bosons.
In unitary gauge we have simply $X = h$,
so \Eq{deVh} can be written in a gauge invariant way as
\[
\de V = \sum_{n \, = \, 3}^\infty \frac{\de\la_n}{n!} X^n.
\eql{deVX}
\]
The variable $X$ is not a local operator expanded around $H = 0$,
but it is a sum of local operators when expanded about the physical
VEV, which is what we will do for the rest of the paper.  As mentioned earlier, expanding around the VEV also shows why the HEFT and SMEFT frameworks are equivalent, since both lead to a power series in $2v h + h^2+\vec{G}^2$.

For example, if we assume that the potential \Eq{deVh} contains
only a modification of the $h^3$ term, then the gauge invariant
form contains terms with arbitrarily high powers of the Higgs and
Nambu-Goldstone fields.
A particularly simple class is the ones with 2 powers of the 
Nambu-Goldstone fields \cite{Falkowski:2019tft}:
\[
\eql{Vcubiconly}
\de V \supset \frac{\de\la_3}{4v} h^2 \vec{G}^2 \sum_{m \, = \, 0}^\infty
\left( -\frac{h}{v} \right)^m.
\]
These give scattering amplitudes for the unphysical Goldstone fields 
that grow with the center of mass energy.
By the equivalence theorem these  are equal to physical $W_L$ and $Z_L$ 
scattering amplitudes at high center of mass energy, and
therefore lead to a violation of tree-level unitarity at high energies.%

In Appendix A we derive unitarity constraints from non-derivative interactions
involving many fields.  For an interaction of the form 
\[
\scr{L}_\text{int} = 
\frac{\la_n}{n_1!\cdots n_r!} \phi_1^{n_1} \phi_2^{n_2} \cdots \phi_r^{n_r},
\]
we derive a unitarity bound on the center of mass energy $E_k$, 
\[
\!\!\!\!\!\!\!
E_k \leq 4\pi \left(\frac{8\pi \displaystyle  \prod_{i \, = \, 1}^r  n_i!}{\displaystyle\la_n \prod_{i \, = \, 1}^r \mat{ n_i \\ k_i}}
\right)^{1/(n-4)} 
\!\!\!
\left(\frac{(k-1)!(k-2)! (n-k-1)! (n-k-2)!}{\displaystyle\prod_{i \, = \, 1}^r k_i!(n_i-k_i)!}\right)^{1/(2n-8)}
\eql{unitaritybound}
\]
for the process $\phi_1^{k_1} \cdots \phi_r^{k_r} \leftrightarrow \phi_1^{n_1-k_1} \cdots \phi_r^{n_r-k_r}$ where we've defined $n\equiv n_1 + \cdots + n_r, k \equiv k_1 + \cdots + k_r$.  This general formula automatically takes into account combinatorial factors from Bose statistics.  
The best bounds come from
processes where the fields are equally distributed between the initial and final 
state, $k_i = n_i/2$ for even $n_i$.  
For additional details, see Appendix A.

\begin{figure*}[t]
\begin{center}
\begin{minipage}{5.75in}
\begin{center}
\includegraphics[width=0.75\linewidth]{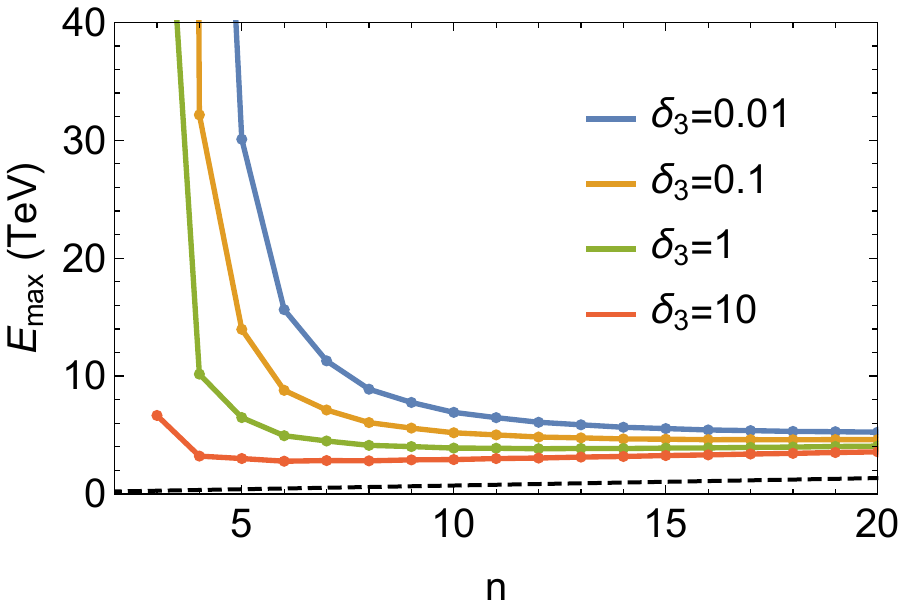}
\end{center}
\small
\caption{
The unitarity violating scale for the interaction $Z_L^2 h^n$ as a function of $n$ 
for different values of $\delta_3$. 
The dashed line shows the threshold energy $\frac{n}{2}m_h + m_Z$, which compared with the 
unitarity violating scale shows that $E_\text{max}$ is large enough to justify the use of the 
equivalence theorem and massless phase space. 
The best limits are
$(\de_3, n, E_\text{max}/\TeV) = (0.01, 22, 5.2),\, (0.1, 18, 4.6),\, (1, 12, 3.8),\, (10, 6, 2.8)$.
\label{fig:ZLZLhtothem}}
\end{minipage}
\end{center}
\end{figure*}

Using the interactions in \Eq{Vcubiconly} we can now give a unitarity bound
for the process $Z_L h^{n/2} \to Z_L h^{n/2}$ for even $n$ 
($Z_L h^{(n-1)/2} \to Z_L h^{(n+1)/2}$ for odd $n$)
in terms of the fractional modification of the Higgs trilinear interaction
\[
\eql{delta3}
\de_3 = \frac{\de\la_3}{\la_3^{\text{(SM)}}} = \frac{v \de\la_3}{3 m_h^2}.
\]
The result is shown in Fig.~\ref{fig:ZLZLhtothem}.
The unitarity bounds are strongest for $n \sim 6$ to 22.
Note that we require $E_\text{max} \gg \frac{n}{2} m_h + m_Z$ to justify 
the use of the equivalence theorem and the use of massless phase space,
but this is well satisfied as can be seen in Fig.~\ref{fig:ZLZLhtothem}.
The bounds are quite strong.
For example, for $\de_3 \sim 1$ the theory violates unitarity near 4 TeV,
and even for $\de_3 \sim 0.01$ the unitarity violation scale is
near 5~TeV.  
The bound is weakly dependent on $\de_3$ because for large multiplicity
the dependence on the coupling is reduced, as can be seen from 
\Eq{unitaritybound}.
These unitarity scales are low enough that they are plausibly within reach of 
experimental searches at the LHC and future colliders.

\begin{figure*}[t]
\begin{center}
\begin{minipage}{5.75in}
\begin{center}
\includegraphics[width=\linewidth]{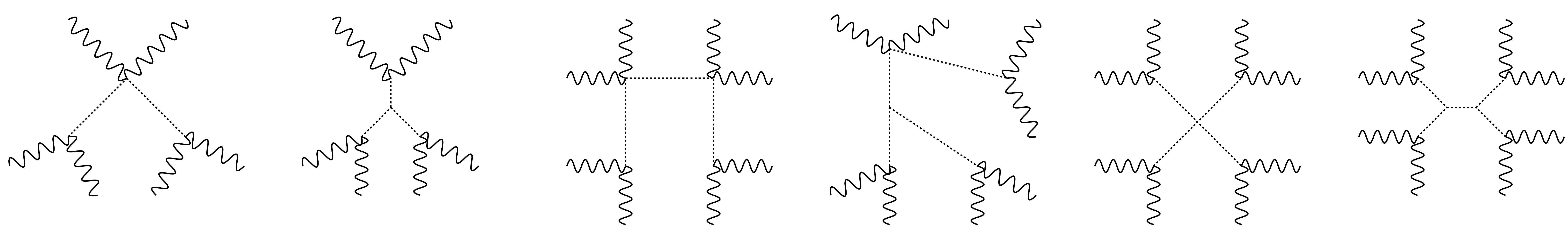}
\end{center}
\small\caption{Representative Feynman diagrams for the $Z_L^6$ and $Z_L^8$ processes
in unitary gauge, 
demonstrating the dependence on the trilinear and quartic Higgs interactions.
\label{fig:Zdiagrams}}
\end{minipage}
\end{center}
\end{figure*}

The existence of tree-level unitarity violating processes involving 
$W_L$ and $Z_L$ can also be understood directly from
tree-level Feynman diagrams, without the formalism introduced above.
The point is that some of the tree-level diagrams that contribute to
these processes involve the Higgs cubic coupling (see for example, Fig.~\ref{fig:Zdiagrams}).
In the SM, there are cancellations in the high-energy behavior of
the amplitude that depend on the Higgs cubic coupling having the
SM value.
When the cubic coupling deviates from the SM value, this cancellation
is absent and the amplitude has harder high-energy behavior.
The utility of the formalism discussed here is that it makes it 
easy to identify the leading high energy behavior in amplitudes
involving many initial and final state particles.

Modifying only the $h^3$ term may appear 
to be a reasonable phenomenological model, but we have seen that
it makes a dramatic prediction of 
tree-level unitarity violation at low energy scales.  
Before concluding that a modification of the cubic coupling
implies new physics at such low scales, we must determine whether
this conclusion is robust.
In fact, it is easy to see that it is not, because there can be cancellations
coming from higher order terms of the form $X^n$ in \Eq{deVX}.
For example, if the modification consists of only the SMEFT
operator $Y^3$, we cannot have any terms higher than $h^6$, since
$Y = vh + \sfrac 12 (h^2 + \vec{G}^2)$.
(In fact, it is easily checked that
$Y = vX + \sfrac 12 X^2$ exactly.)
A cubic modification alone on the other hand, involves an infinite series in $Y$, 
\[
X^3 &= \bigl(\sqrt{v^2+2 Y}-v \bigr)^3 
=  \frac{Y^3}{v^3} - \frac 32 \frac{Y^4}{v^5}
+ \frac{9}{4} \frac{Y^5}{v^7}
- \frac{7}{2} \frac{Y^6}{v^9}
+ \cdots
\eql{cubicalone}
\]
whose coefficients 
in units of $v$
do not fall off for higher powers,
and it is therefore not surprising that this predicts
high multiplicity processes with low scales of unitarity violation.  
These examples show that
the existence of contact interaction terms with many Higgs bosons,
which were the origin of the strong unitarity bounds derived above,
is not a model-independent consequence of a deviation in the $h^3$
coupling.

The lesson is simply that we must consider the most general possible
modification of the Higgs potential in order to draw robust conclusions
about the high-energy behavior of the theory.
To see that there are growing amplitudes for a general modification,
we expand the potential \Eq{deVX} in powers of $h$ and $\vec{G}$.
Powers of $X$ have the structure (see \Eq{theX})
\begin{subequations}
\eql{Xstructure}
\[
X^3 &\sim h^3 + \vec{G}^2(h^2 + h^3 + \cdots)
+ \vec{G}^4(h + h^2 + \cdots)
+ \vec{G}^6(1 + h + \cdots)
\nn
&\qquad\ \gap{}
+ \vec{G}^8(1 + h + \cdots) + \vec{G}^{10}(1 + h + \cdots) + \cdots,
\\
X^4 &\sim h^4 + \vec{G}^2(h^3 + h^4 + \cdots)
+ \vec{G}^4 (h^2 + h^3 + \cdots)
+ \vec{G}^6 (h + h^2 + \cdots)
\nn
&\qquad\ \gap{}
+ \vec{G}^8(1 + h + \cdots) + \vec{G}^{10}(1 + h + \cdots) +\cdots,
\\
X^5 &\sim h^5 + \vec{G}^2(h^4 + h^5 + \cdots)
+ \vec{G}^4 (h^3 + h^4 + \cdots)
+ \vec{G}^6 (h^2 + h + \cdots)
\nn
&\qquad\ \gap{}
+ \vec{G}^8(h + h^2 + \cdots) 
+ \vec{G}^{10}(1 + h + \cdots)
+ \cdots,
\]
\end{subequations}
where we set $v = 1$.
From this we see that the potential terms
\[
\eql{mod3}
V \supset
\frac{m_h^2}{4v^2}(1+3\de_3)\, \vec{G}^2 h^2
+
\frac{3m_h^2}{8v^3}\de_3\, \vec{G}^4 h
+
\frac{m_h^2}{16v^4}\de_3\, \vec{G}^6,
\]
arise only from the $X^3$ term,
and are therefore determined by the deviation of the Higgs cubic
term in the potential independently of the rest of the Higgs potential.
(Note that the interaction $\vec{G}^2 h^2$ is already present in the SM 
Higgs potential.)

To robustly determine the scale of tree-level unitarity violation implied by a
modification of the Higgs cubic, we consider tree-level amplitudes 
of the fields $h$ and $\vec{G}$ that 
get contributions from
the interaction terms
\Eq{mod3}.
We will see below that the strongest bound comes from 3-to-3 processes
such as $Z_L^3 \leftrightarrow Z_L^3$.
We will compute this using the equivalence theorem below, but we first
consider the calculation in unitary gauge.
The tree-level amplitude gets contributions from diagrams like the first two diagrams
of Fig.~\ref{fig:Zdiagrams}.
The first diagram represents 45 different terms obtained by permutations of
external legs and vertices, while the second represents 15.  
The complexity of this calculation explains why early work on these processes 
\cite{Belyaev:2012bm} could only be analytically calculated with all of the 
particle momenta restricted to a common plane and motivates using the much 
simpler equivalence theorem calculation.  
At high energies, there are diagrams that are independent of $E$
at high energies, but for the SM value of the Higgs cubic these terms cancel
and the amplitude goes as $1/E^2$ at high energy, as required by unitarity.   By summing all of these together, one could verify that if  the Higgs trilinear interaction is the Standard Model value, the diagrams cancel to achieve the required energy behavior, $1/E^2$, for a unitary six point amplitude.  
However, if the trilinear is nonstandard, the sum is a constant at high energies that is proportional to $\delta_3$.  

These results for the six $Z_L$ process are much simpler to see using the equivalence theorem.  Our potential interactions for the Goldstones do not involve derivatives, so the amplitude's energy dependence comes simply from propagators.  Thus, the leading energy dependence is constant and comes from the $\vec{G}^6$ contact interaction, which is proportional to $\delta_3$.  If the Higgs trilinear has the standard value, then there is no six point contact interaction and the amplitude falls off as $1/E^2$ from diagrams with a single propagator.          
If we now calculate the leading piece, using the results from the Appendix, we obtain the unitarity
bound
\[
E_\text{max} \lsim \frac{16\TeV}{|\de_3|^{1/2}}.
\]
Bounds for other processes are given in Table~\ref{tbl:Uprocesses}. The strongest five particle process that depends only on the trilinear modification is $h Z_L^2 \leftrightarrow Z_L^2$, with the bound 
\[
E_\text{max} \lsim \frac{94\TeV}{|\de_3|}
\]
which gives a stronger bound only for $|\de_3| > 35$, which violates the current LHC constraints on the trilinear.

Optimized bounds for the $\vec{G}^6$ interaction can be found by diagonalizing the transition matrix element.  Using custodial $SU(2)$ symmetry, we can categorize the allowed scattering channels.  For 3 $G$'s to 3 $G$'s scattering, there is both a $I=1$ and a $I=3$ channel.  As detailed in the Appendix, the $I=1$ channel sets the best limit, with   
\[
\eql{theEmax}
E_\text{max} \lsim \frac{13.4\TeV}{|\de_3|^{1/2}}.
\]
A similar analysis can be done for the $h\vec{G}^4$ interaction.  Here the allowed channels are $I=0, 1,$ and 2.  The best bound comes from the $I=0$ channel, with the bound
\[
E_\text{max} \lsim \frac{57.4 \TeV}{|\de_3|}.
\]
These bounds improve a bit upon the channels earlier explored, giving a $\sim 20-40\%$ reduction in the energy scale for unitarity violation.

\begin{table}
\begin{center}
\begin{tabular}{|c|c|}
\hline
Process & Unitarity Violating Scale \\ 
\hline\hline
$h^2 Z_L \leftrightarrow h Z_L$ & ${66.7\rm{\; TeV}}/{|\delta_3-\frac{1}{3}\de_4|}$ \\ \hline
$h Z_L^2 \leftrightarrow  Z_L^2$ & ${94.2\rm{\; TeV}}/{|\delta_3|}$ \\\hline
$h W_L Z_L \leftrightarrow W_L Z_L$ & ${141\rm{\; TeV}}/{|\delta_3|}$ \\ 
\hline \hline
$h Z_L^2 \leftrightarrow h Z_L^2$ & ${9.1\rm{\; TeV}}/{\sqrt{|\delta_3-\frac{1}{5}\de_4|}}$ \\ \hline
$h W_L Z_L \leftrightarrow h W_L Z_L$ & ${11.1\rm{\; TeV}}/{\sqrt{|\delta_3-\frac{1}{5}\de_4|}}$\\ \hline
$Z_L^3 \leftrightarrow Z_L^3$ & ${15.7\rm{\; TeV}}/{\sqrt{|\delta_3}|}$\\ \hline
$Z_L^2 W_L \leftrightarrow Z_L^2 W_L$ & ${20.4\rm{\; TeV}}/{\sqrt{|\delta_3|}}$ \\ 
\hline \hline
$h Z_L^3 \leftrightarrow Z_L^3$ & ${6.8\rm{\; TeV}}/{|\delta_3-\frac{1}{6}\de_4|^{\frac{1}{3}}}$  \\ \hline
$h Z_L^2 W_L \leftrightarrow Z_L^2W_L$ & ${8.0\rm{\; TeV}}/{|\delta_3-\frac{1}{6}\de_4|^{\frac{1}{3}}}$  \\ 
\hline\hline
$Z_L^4 \leftrightarrow Z_L^4$ & ${6.1\rm{\; TeV}}/{|\delta_3-\frac{1}{6}\de_4|^{\frac{1}{4}}}$ \\ \hline
\end{tabular}
\begin{minipage}{5.75in}
\small\caption{\label{tbl:Uprocesses} Unitarity violating amplitudes that only depend on the trilinear and quartic Higgs modifications.}
\end{minipage}
\end{center}
\end{table}

Let us consider what happens if we also include the effects of the quartic
interaction.
From \Eq{Xstructure}, we see that the new terms which depend only on the
$h^3$ and $h^4$ modification are
\[
V &\supset \frac{m_h^2}{8v^2}(1+\de_4)\, h^4
+ 
\frac{m_h^2}{4v^3}(\de_4-3\de_3)\, h^3 \vec{G}^2
+
\frac{3m_h^2}{16v^4}(\de_4-5\de_3)\, h^2 \vec{G}^4
\nn
&\qquad{}
+ \frac{m_h^2}{16v^5}(\de_4-6\de_3)\, h \vec{G}^6
+\frac{m_h^2}{128v^6}(\de_4-6\de_3)\, \vec{G}^8.
\eql{mod4}
\]
These can give stronger unitarity bounds, depending on the value of the
deviation in the Higgs quartic interaction, see Table~\ref{tbl:Uprocesses}.
For example, the process $Z_L^4 \leftrightarrow Z_L^4$, which would normally require evaluation of several diagrams as shown in Fig.~\ref{fig:Zdiagrams}, can be easily analyzed with the equivalence theorem to give a unitarity bound
\[
\eql{G8}
E \lsim \frac{6.1\TeV}{\bigl|\de_3 - \sfrac 16 \de_4\bigr|^{\frac 14}},
\]
where we define the fractional quartic coupling deviation
\[
\de_4 = \frac{\de\la_4}{\la_4^{\text{(SM)}}} = \frac{v^2 \de\la_4}{3 m_h^2}.
\]
\Eq{G8} is the unitarity bound that arises from a single insertion of the
$\vec{G}^8$ contact term that arises from the $X^3$ and $X^4$ terms in 
the effective Higgs potential.  
There are also unitarity-violating contributions to the 
$Z_L^4 \leftrightarrow Z_L^4$ amplitude from tree-level diagrams
with internal lines, but these are parametrically smaller for
$\de_3 \sim \de_4 \lesssim 1$.
For example, there is a contribution with two insertions of the 
$h\vec{G}^4$ coupling with a Higgs propagator, which gives a
contribution to the amplitude of order
\[
\de\scr{M}(Z_L^4 \to Z_L^4)
\sim \left( \frac{\de_3 m_h^2}{v^3} \right)^2 \frac{1}{E^2}.
\]
which is parametrically small compared to the contribution that gives the 
bound \Eq{G8}:
\[
\scr{M}(Z_L^4 \to Z_L^4) \sim \frac{(\de_4 - 6 \de_3) m_h^2}{v^6}.
\]
As noted earlier, it is difficult to experimentally constrain 
the Higgs quartic interaction even at future colliders,
so it is unlikely that one can use \Eq{G8} to give an experimental
estimate of the scale of new physics.
This is unfortunate, since for generic values of $\de_3, \de_4 \lsim 1$, 
the bound \Eq{G8} is stronger than the model-independent bound \Eq{theEmax}.
However, this is not completely model-independent, since
the bound \Eq{G8} disappears for $\de_4 = 6 \de_3$.
In fact, this special choice corresponds to having only a $Y^3$ coupling,
which is natural from a UV point of view;
we will discuss this point further below.
It also clarifies the large difference between the model-independent bound
\Eq{theEmax}, and the stronger bounds obtained when we assumed that
only the $h^3$ term was modified (see Fig.~\ref{fig:ZLZLhtothem}).
Those stronger bounds come from processes with high multiplicity,
and these can be cancelled if we allow arbitrary conspiracies among
couplings.

So far we have only considered Higgs potential interactions.
For completeness, we must also consider the effect of including
derivative interactions in our model-independent bound.
For example, we can consider the term
\[
\de\scr{L} = \frac{1}{f} X (\d X)^2 \supset \frac{1}{f} h (\d h)^2.
\eql{derivativecubic}
\]
which gives a momentum dependent contribution to the Higgs three point function.
This mimics a Higgs trilinear $\de_3 \sim v/f$ near the threshold for Higgs pair production.
However, because of the extra derivatives, this 
gives rise to unitarity violating processes that grow faster with
energy than the potential modifications.
For example, $h h \to Z_L Z_L$ has a matrix element 
$\scr{M} \sim E^2/(f\, v)\sim \de_3 E^2/v^2 $,
with a unitarity bound $E\lsim \sqrt{16\pi v^2/|\de_3|} \sim 2 \TeV/\sqrt{|\de_3|}$.  
We see that attempting to explain a large deviation in the Higgs trilinear
coupling with derivative couplings results in a unitarity violating scale that is
\emph{lower} than the model-independent bound \Eq{theEmax}, 
because they give rise to amplitudes that grow faster at high energies.

As a note of caution, we warn that
care must be taken when using equations of motion 
(or equivalently nonlinear field redefinitions) to simplify the Lagrangian
\cite{Georgi:1991ch, Arzt:1993gz}. 
For example,
the nonlinear 
redefinition
\[
\eql{Xredfn}
X \to X - \frac{1}{2f} X^2 
\]
(equivalent to redefining $h$)
eliminates the coupling in \Eq{derivativecubic} 
at linear order in the deviation from the SM.
However, this redefinition also changes the SM part of the Lagrangian,
which will now have interactions that give rise to unitarity violation.
For example, the redefinition \Eq{Xredfn} modifies the $hhZZ$ coupling,
and we find an amplitude for $h h \to Z_L Z_L$ that grows as $E^2$.  
A closely related example is the case of the
dimension-6 operator
\[
\de\scr{L} = \frac{1}{\Lambda^2} (\d |H|^2)^2 \supset \frac{1}{\Lambda^2} (v+h)^2 \d h^2.
\]   
Note that this modifies the Higgs kinetic term as well as generating the 
coupling \Eq{derivativecubic} with $f = \La^2/2v$.
As shown in \cite{Giudice:2007fh}, a nonlinear field transformation 
\[
h\to h-\frac{v^2}{\Lambda^2}\left(h+\frac{h^2}{v}+\frac{h^3}{3v^2}\right)
\]
removes to linear order the derivative self-interactions but leads to modifications 
of the non-derivative Higgs self-interactions (including the Higgs trilinear) 
and Higgs couplings to $W,Z$ and top quark.  
After the transformation the matrix elements for the unitarity violating processes 
are unchanged, but are much easier to see directly by using the original 
form of the derivative operator.

As a final example, we consider the coupling
\[
\eql{hkineticmod}
\de\scr{L} = \sfrac 12 \de Z_h (\d X)^2 = \sfrac 12 \de Z_h (\d h)^2 + O(\vec{G}^2),
\]
which changes the normalization of the $h$ kinetic term, and therefore the
physical Higgs cubic coupling by $\de_3 \sim \de Z_h$.
It also changes the Higgs couplings to all other SM particles.
Because of this there are many additional processes that violate unitarity,
but the unitarity violating processes that arise from the modified Higgs
cubic are still present, and so these processes cannot change
our model-independent bounds.%
\footnote{The couplings of the Higgs to the other SM fields such as the
$W$ and $Z$ bosons are more accurately known to agree with the SM than the
Higgs self-coupling, so it would seem that we require $\de Z_h \lsim 0.1$.
However, we can artificially cancel these deviations by adding terms such
as 
\[
\de\scr{L} = \bigl( \sfrac 12 m_Z^2 Z^\mu Z_\mu 
+ m_W^2 W^{+\mu} W^-_\mu \bigr)
\frac{h}{v}.
\]
If we do this, we also cancel the growing amplitudes that involve these
additional couplings, but the bound from the Higgs trilinear deviation
still applies.}

To summarize the results of this section,
a ``generic'' modification of the low-energy Higgs potential 
gives rise to unitarity violation at a few TeV.
This unitarity violation arises in processes involving many Higgs and gauge 
particles and $W$ and $Z$ bosons, and these can be canceled if the
parameters of the low-energy Higgs potential obey special relations.
Allowing for these cancellations, there is still a model-independent bound
that depends only on the deviation in the Higgs cubic, but the scale of
unitarity violation is much higher ($\sim 13$~TeV).
This scale is potentially accessible to a future 100~TeV $pp$ collider,
although it will be more challenging.

\section{Models with a Nonstandard Higgs Trilinear Coupling}
\label{sec:TrilinearModels}
The results of the previous section raise the question of whether the model-independent
bound is too conservative.
After all, from the bottom-up point of view it appears to require a large number
of conspiracies among low-energy parameters to avoid the much stronger
bound for the ``generic'' modification of the Higgs potential.
However, we will now show that it is difficult to construct a UV theory with
strongly coupled interactions between many Higgs and gauge bosons, as
would be required to saturate the ``generic'' bound.
On the other hand, the model-independent
bound can be saturated in a UV model with Higgs compositeness at the unitarity
violating scale.
The model that saturates the bound requires fine-tuning of 2 parameters
(compared to 1 parameter in the SM itself), but it gives an existence proof
that the model-independent bound can be saturated in a sensible UV theory.
If we reduce the tuning in the model, we find that the scale of new physics
goes below the unitarity bound, suggesting that the model-independent bound
may be too conservative.

We now construct a UV theory that gives rise to the desired modification of the Higgs
trilinear, and where the new physical appears at the model-independent unitarity
bound \Eq{theEmax}.
In such a theory, the couplings of the Higgs must get strong
at a scale $M = 13.4~\text{TeV}/|\de_3|^{1/2}$, where the new particles enter to 
unitarize the theory.
Because we want the \emph{non-derivative} interactions in the Higgs potential
to become strong at the scale $M$, we do not consider models where the
Higgs is a pseudo Nambu-Goldstone boson.
If we assume that the strength of the interactions of the Higgs at the scale
$M$ is given by a dimensionless coupling $g_*$, then the effective Lagrangian
below the scale $M$ is given by 
\[
\de\scr{L}_\text{eff} &= \frac{M^4}{g_*^2} F\left( \frac{g_* H}{M}, \frac{\d}{M}, \ldots\right)
\nn
&\sim D^\mu H^\dagger D_\mu H 
+ M^2 H^\dagger H + \frac{g_*^2}{2!} (H^\dagger H)^2
+ \frac{g_*^4}{3! M^2} (H^\dagger H)^3
 + \frac{g_*^2}{2! M^2} \bigl( \d_\mu |H|^2 \bigr)^2
 + \cdots\,.
\eql{Hcomp}
\]
This is a modification of the SILH power counting
to the case where the Higgs is not a pseudo-Nambu-Goldstone boson 
\cite{Giudice:2007fh,Azatov:2015oxa}.
For $g_* \sim 4\pi / \sqrt{N}$ 
\Eq{Hcomp} reproduces ``\naive dimensional analysis'' (NDA)
for large-$N$ theories
\cite{Manohar:1983md, Georgi:1986kr},
which works reasonably well in estimating the size of terms in the effective
chiral Lagrangian of QCD, as well as in calculable strongly-coupled
SUSY theories \cite{Luty:1997fk,Cohen:1997rt,Luty:1999qc}.
The second line in \Eq{Hcomp} should be taken as a rough approximation.
For example, the factors of $1/n!$ multiplying $(H^\dagger H)^n$ can be justified
for large $n$, but may be questioned for small values of $n$.
We will assume that $M$ is given by our conservative unitarity bound in \Eq{theEmax}, and fix $g_*$ from the
Higgs trilinear deviation $\de_3$.
We obtain $g_* \sim 6.9$, independent of $\de_3$.
In this model, the effect of higher order terms in the potential at low
energies is suppressed compared to the $(H^\dagger H)^3$ term by powers
of
\[
\left(\frac{g_* v}{M}\right)^2 \sim \frac{1}{60} \, |\de_3|,
\]
justifying the use of the low-energy expansion in this model.
Also, terms involving derivatives are suppressed compared to those without
derivatives by powers of $1/g_*$.
For example, the operator $\d^\mu (H^\dagger H) \d_\mu (H^\dagger H)$ contributes
a contribution to the Higgs trilinear deviation
$\de(\de_3)/\de_3 \sim m_h^2 / (g_*^2 v^2) \sim 0.005$. 
The difficulty with a model of this kind is that the $H^\dagger H$ and $(H^\dagger H)^2$ 
terms are much too large compared to the Standard Model values.  
There is no symmetry difference between the various powers of $H^\dagger H$,
in \Eq{Hcomp}, and it appears that the only way to get agreement with the 
SM is to fine-tune the $H^\dagger H$ and $(H^\dagger H)^2$ terms 
to be small.
\footnote{There is no sign constraint on the coefficients of the
 $H^\dagger H$ and $(H^\dagger H)^2$ terms, so we expect that there
are fine-tuned models where their coefficients can be anomalously small
for special choices of parameters.
For example, a negative $(H^\dagger H)^2$ term can arise from the tree-level
exchange of a massive singlet scalar field  $S$ with a $S H^\dagger H$ coupling.}
The overall tuning is %
the product of the tuning of the two terms, and is given by
\[
\text{tuning} \sim \frac{|m_H^2|}{M^2} \frac{\la_H}{g_*^2}
\sim \frac{|\de_3|}{5 \times 10^6}.
\]
The tuning gets worse for small $\de_3$, because the scale of new physics
required to get the deviation of the Higgs cubic becomes larger.
The tuning of the $H^\dagger H$ term could be explained by anthropic arguments
\cite{Agrawal:1997gf},
but there is no 
anthropic reason
for the tuning of the $(H^\dagger H)^2$ term, 
so such a model still has an unexplained tuning of order $1/500$. 
In other words, this model does not naturally account for the fact that the
Higgs appears to be a weakly-coupled particle at low energies. 
A model of this kind
is not an attractive paradigm for physics beyond the SM, but it 
does provide
an existence proof for models that saturate the model-independent unitarity bound.  

If we consider more natural UV models, the scale of new physics is below the
model-independent unitarity bound.
For example, we can consider a model of the type \Eq{Hcomp}, but with a smaller
value of $g_*$.
Such a model requires a lower value of $M$ to explain a given deviation $\de_3$,
simultaneously making the model more natural while lowering the scale
of new physics.
For example, for $g_* \sim 1$ the UV physics is weakly coupled,
and the scale of new physics is given by
\[
M \sim \sqrt{\frac{1}{c_3}}= \sqrt{\frac{v^4}{3m_h^2\de_3}} 
= \frac{280\GeV}{\sqrt{|\de_3|}}.
\]
Such a model therefore requires new physics at the electroweak scale,
and this kind of new physics is
strongly constrained by direct searches, electroweak precision tests,
and also obtaining the observed Higgs mass.
In fact, as reviewed in \cite{Gupta:2013zza}, in many beyond the Standard Model 
frameworks ({\it e.g.}~supersymmetry, composite Higgs) it is difficult to have
modifications of the Higgs trilinear larger than $10-20\%$ due to these constraints.
A natural framework for new physics that allows somewhat larger deviations is
induced electroweak symmetry breaking 
\cite{Azatov:2011ht,Azatov:2011ps,Galloway:2013dma,Chang:2014ida}, 
which also requires new physics below the TeV scale.

Are there reasonable UV models with new particles at the ``generic''
unitarity violating scale $E_\text{max} \sim 5$~TeV 
(see Fig.~\ref{fig:ZLZLhtothem}), 
with the correct low energy expansion?
This bound arises from processes involving many particles, so the
basic requirement is that higher powers of $H$ are suppressed by powers
of $v$ with order-1 coefficients, for example
\[
\de V \sim \de_3 m_h^2 v^2  \sum_{n \, = \, 3}^\infty
a_n \left( \frac{H^\dagger H}{v^2} \right)^n,
\qquad
a_n \sim 1.
\]
We want to reproduce this in a UV model where the Higgs is a composite 
particle with strong interactions at the scale $M$.
As previously noted, we want the non-derivative terms to violate
unitarity, so we do not assume that the Higgs is a pseudo Nambu-Goldstone
boson.
We then expect the potential to be given by the power counting \Eq{Hcomp},
which requires $M \sim g_* v$.
For $M \sim 5$~TeV, this gives $g_* \sim 20$, which is even stronger than the
strongest coupling one would expect based on considerations of unitarity
or NDA.
More problematically, matching the prefactor in the potential requires
\[
\de_3 \sim \frac{M^2}{3 m_h^2} \sim 500.
\]
Thus, to saturate the generic bound, the Higgs VEV, mass and trilinear must 
all be tuned
to be consistent with current constraints, requiring a total tuning of $\sim 10^{-10}$.  
 
Our conclusion that it is difficult to construct a UV model that 
generates a ``generic'' deviation in the Higgs potential.
On the other hand, we have shown that if the Higgs is composite
it is possible to saturate the weaker model-independent bound.
However, even this model is very fine-tuned, and thus we expect the scale of new
physics to be below the model-independent bound.
For example, 
induced electroweak symmetry breaking is an existence proof of a 
class of models that have large deviations in the Higgs trilinear, while
giving a natural explanation of the successes of the SM.
We believe that these considerations %
only
strengthen the 
motivation for the measurement of the Higgs cubic.

\section{Conclusions} \label{sec:conclusions}
We have considered the scales of unitarity violation in a theory where the
low-energy Higgs potential is modified from the Standard Model prediction.
The Standard Model predicts precise cancellations among different diagrams
to guarantee good high energy behavior, and any deviation from the 
Standard Model predictions for couplings will upset this behavior and
lead to the breakdown of perturbation theory at high energies.
This is a classic argument that was used to predict the existence of new
physics below the TeV scale in the theory without a Higgs sector,
providing a ``no lose'' theorem for the LHC.

We extended this argument to a theory with a Higgs, but with modifications
of the Higgs potential.
Using the equivalence theorem, we can determine which processes involving
longitudinal $W$'s and $Z$'s and Higgs particles violate unitarity, and easily
compute their high-energy behavior.
We have shown that generic modifications of the Higgs trilinear coupling
lead to the theory breaking down near 5~TeV, nearly independently of the
size of the deviation.
This scale is tantalizingly close to the energy scale currently being
probed by the LHC.
However, the bad high energy behavior can be canceled by deviations
in higher order terms in the Higgs potential, and is therefore not
model-independent.
We find that there is a completely model-independent bound on the
scale of new physics that depends only on the modification of the Higgs
trilinear coupling at low energies:
the theory must break down at a scale $\lsim 13\TeV/|\de_3|$, where $\de_3$ is the fractional  
modification of the Higgs trilinear coupling.
This means that measurements of the Higgs trilinear directly point to a new
UV energy scale where new physics must appear, giving additional motivation
for these searches.
If any deviation is observed, it would provide a target for future high
energy colliders designed to explore this higher energy scale.

The growing amplitudes discussed in this work motivate experimental searches
in these channels.
Much of the existing work on searching for new high energy physics in
electroweak final states has focused on low multiplicity final states,
such as $Zh$ \cite{Banerjee:2018bio}, $tW$ \cite{Dror:2015nkp}, 
and $(tZ,th) + \text{jet}$ \cite{Degrande:2018fog}.
A recent paper \cite{Henning:2018kys} has considered amplitudes whose energy 
growth results from Higgs coupling modifications.  
One of their analyses probed $VV \to WWh$ ($V = W$ or $Z$) using vector boson fusion,
and found sensitivity to $|\de_3|\gsim 5$ at the high luminosity LHC.
Given our analysis, it would be interesting to explore unitarity systematically
for other processes such as $VV \to VVVV$.

\section*{Acknowledgements}
We thank A.~Falkowski and R.~Rattazzi for discussions about their closely
related paper \cite{Falkowski:2019tft}.   
We especially thank E.~Salvioni for extensive discussions and comments on the draft. 
SC is supported in part by the U.S. Department of Energy under grant DE-SC0011640.
ML is supported in part by the U.S. Department of Energy under grant DE-SC-0009999.

\appendix{Appendix A: Unitarity Bound on Potential Interactions\label{app:generalunitarity}}
We are interested in the bounds placed on tree-level scattering amplitudes by
unitarity.
The idea of these unitarity constraints is very simple.
We write the $S$ matrix as
\[
S = \id + i T,
\]
where the identity contribution represents the free propagation of
particles without interactions, and the transition matrix
$T$ describes interactions.
Unitarity of the $S$ matrix implies that if $\ket{i}$ and $\ket{f}$ are unit
normalized states we have
\[
\bigl|\bra{f} S \ket{i}\bigr| \le 1
\quad
\text{for all }i, j.
\]
For $\ket{f} \ne \ket{i}$, this implies
\[
\bigl|\bra{f} T \ket{i}\bigr| \le 1.
\]

Plane-wave states are not unit normalized, but we can define normalized states
using the partial wave expansion.
More generally, we can label the initial and final states by the total
4-momentum $P^\mu$, and we assume that the additional quantum numbers
$\al$ required to specify the state are discrete, so that the states are
normalized to
\[
\eql{discretenorm}
\braket{P', \al'}{P, \al} =  (2\pi)^4 \de^4(P - P') \de_{\al\al'}.
\]
For example, in the partial wave expansion of a state of 2 scalar particles,
we can take $\al$ to consist of the relative angular momentum quantum numbers
$\ell$ and $m$.
Defining the Lorentz invariant amplitude
\[
\eql{dicreteamp}
\bra{P', \al'} T \ket{P, \al} = (2\pi)^4 \de^4(P - P') \scr{M}_{\al'\al},
\]
we have
\[
\eql{dicreteS}
\bra{P_f, \be} S \ket{P_i, \al}
= (2\pi)^4 \de^4(P_f - P_i) S_{\be\al},
\]
with
\[
S_{\be\al} = \de_{\be\al} + i \scr{M}_{\be\al}.
\]
Unitarity of the $S$-matrix implies that the matrix $S_{\be\al}$ is
unitary, and the same logic as above implies that
\[
|\scr{M}_{\be\al}| \le 1
\]
for $\beta \neq \alpha$.  For a bound when $\beta=\alpha$, consider
\[
1= \delta_{\al\al} = \sum_\gamma S^*_{\gamma \al}S_{\gamma\al} = 1 -2\, \text{Im}\; \scr{M}_{\al \al} + \sum_\gamma |\scr{M}_{\gamma \alpha}|^2
\]
which gives
\[
2\, \rm{Im}\, \scr{M}_{\alpha\alpha} = \sum_{\gamma} |\scr{M}_{\gamma\alpha}|^2\geq  |\scr{M}_{\alpha\alpha}|^2 = |\rm{Re}\, \scr{M}_{\alpha\alpha}|^2+|\rm{Im}\, \scr{M}_{\alpha\alpha}|^2.
\]
Completing the square shows that $1\geq |\rm{Re}\, \scr{M}_{\alpha\alpha}|^2+|\rm{Im}\, \scr{M}_{\alpha\alpha}-1|^2$.  This implies the bounds
\[
|\rm{Re}\, \scr{M}_{\alpha\alpha}| \le 1, \quad 0\leq \rm{Im}\, \scr{M}_{\alpha\alpha} \leq 2.
\] 
Since $\scr{M}_{\alpha\alpha}$ is real at tree level, this implies that $|\scr{M}_{\alpha\beta}| \leq 1$ is true at tree level for all states $\alpha, \beta$.

Let us apply these ideas to obtain the unitarity bound on an effective
interaction term of the form
\[
\eql{Lint}
\scr{L}_\text{int} = 
\frac{\la_n}{n_1!\cdots n_r!} \phi_1^{n_1} \phi_2^{n_2} \cdots \phi_r^{n_r},
\]
where $\phi_i$ are independent, real scalar fields
and $n_i$ are positive integers.
We define the normalized states
\[
\eql{optstates}
\ket{P, k_1,\dots, k_r} = C_{k_1\cdots k_r} \int d^4 x\, 
e^{i P\cdot x} 
\prod_{i \, = \, 1}^r \left[\phi_i^{(-)}(x) \right]^{k_i} \ket{0},
\]
where the $k_i$ are non-negative integers that play the role of the discrete label $\al$ 
in \Eq{discretenorm}, and
$\phi^{(-)}_i$ is the part of the (interaction picture) field $\phi_i$ that 
contains a creation operator.
The idea behind the states in \Eq{optstates}
is that they have the largest overlap with the
interaction \Eq{Lint}, and will therefore give the strongest unitarity bounds.
The normalization \Eq{discretenorm} then fixes
\[
\frac{1}{|C_{k_1 \cdots k_r}|^2} = \frac{1}{(k-1)!(k-2)!} \frac{1}{8\pi} 
\left( \prod_{i \, = \, 1}^r k_i ! \right)
\left(\frac{E}{4\pi}\right)^{2k-4}
\]
where $k =  k_1 + \cdots + k_r$ and we have assumed all the particles are massless. 
Working out the matrix element for the scattering amplitude
with $k_i \to n_i - k_i$, we get 
\[
\scr{M}_{n-k,k} &= \frac{\la_n}{n_1!\cdots n_r!} \frac{1}{C_{n_1-k_1 \cdots n_r-k_r} C_{k_1 \cdots k_r}^*} \prod_{i \, = \, 1}^r \mat{n_i \\ k_i}
\\ 
&= \nonumber  \frac{\la_n}{8\pi \sqrt{(k-1)!(k-2)! (n-k-1)! (n-k-2)!}} \left(\frac{E}{4\pi}\right)^{n-4}
\\ 
&\qquad\quad{} 
\times \prod_{i \, = \, 1}^r \mat{n_i \\ k_i} \frac{\sqrt{k_i! (n_i-k_i)!}}{n_i!}\,,
\]
where $n = n_1 + \cdots + n_r$.
Requiring this to be less than 1 gives the bound
\[
\!\!\!\!\!\!\!
E_k \leq 4\pi \left(\frac{8\pi \displaystyle  \prod_{i \, = \, 1}^r  n_i!}{\displaystyle\la_n \prod_{i \, = \, 1}^r \mat{ n_i \\ k_i}}
\right)^{1/(n-4)} 
\!\!\!
\left(\frac{(k-1)!(k-2)! (n-k-1)! (n-k-2)!}{\displaystyle\prod_{i \, = \, 1}^r k_i!(n_i-k_i)!}\right)^{1/(2n-8)}.
\]
The lowest unitarity limit is when $k_i = \frac 12 n_i$ (assuming all the $n_i$ are even), which improves on the conventionally analyzed $2\to m$ scattering processes.  
This gives 
\[
E_{k \, = \, n/2} \leq 4\pi \left(\frac{8\pi (\frac{n}{2}-1)!(\frac{n}{2}-2)! \prod_{i \, = \, 1}^r (n_i/2)!}
{\displaystyle\la_n }\right)^{1/(n-4)}.
\]
In the large $n$ limit, the bound asymptotically gets worse, although intermediate multiplicities may still give better bounds since it can counteract the  values of $\la_n$ in a given model. 

Using this formula, we can determine the optimal channel (i.e. choice of $k_1, \ldots, k_r$) to get the lowest bound.  For a representative set of five to eight point interactions, in Table~\ref{tbl:bestbounds} we list the optimal energy bound and the channel it can come from (there are multiple choices coming from permutations and swapping of initial and final states). We note that in the above, we have neglected contributions to scattering amplitudes
that involve multiple insertions of the interactions.
These involve diagrams with one or more internal propagator, and
this means that these interactions scale with energy with a power
less than that of the model-dependent terms.

\begin{table}
\begin{center}
\begin{minipage}{5.75in}
\begin{center}
\begin{tabular}{|l|l|c|}
\hline
$(n_1,\dots,n_r)$ & Best $(k_1,\ldots,k_r)$  & $E_\text{max}$ \\ 
\hline\hline
(5) &(2) & $1550/\la_n$\\
(4,1) &(2,0) & $ 893/\la_n$\\
(3,2) &(1,1) & $632/\la_n$\\
(3,1,1) &(1,1,0) & $632/\la_n$\\
(2,2,1) &(1,1,0) & $447/\la_n$\\
(2,1,1,1) &(1,1,0,0) & $ 447/\la_n$\\
\hline\hline
(6) &(3) & $ 218/\sqrt{\la_n}$\\
(5,1) &(2,1) & $ 166/\sqrt{\la_n}$\\
(4,2) &(2,1) & $ 126/\sqrt{\la_n}$\\
(3,3) &(2,1) & $ 126/\sqrt{\la_n}$\\
(4,1,1) &(2,1,0) & $ 126/\sqrt{\la_n}$\\
(3,2,1) &(1,1,1) & $ 106/\sqrt{\la_n}$\\
(2,2,2) &(1,1,1) & $ 89/\sqrt{\la_n}$\\
(3,1,1,1) &(1,1,1,0) & $ 106/\sqrt{\la_n}$\\
(2,2,1,1) &(1,1,1,0) & $ 89/\sqrt{\la_n}$\\
\hline\hline
(7) &(3) & $ 143/\la_n^{1/3}$\\
(6,1) &(3,0) & $ 114/\la_n^{1/3}$\\
(4,2,1) &(2,1,1) & $ 79/\la_n^{1/3}$\\
(2,2,2,1) &(1,1,1,0) & $ 63/\la_n^{1/3}$\\
\hline\hline
(8) &(4) & $ 116/\la_n^{1/4}$\\
(6,2) &(3,1) & $ 82/\la_n^{1/4}$\\
(4,4) &(2,2) & $ 74/\la_n^{1/4}$\\
(4,2,2) &(2,1,1) & $ 62/\la_n^{1/4}$\\
\hline
\end{tabular}
\small\caption{The lowest unitarity violating energy scales
for some five to eight point
interactions of the form  \Eq{Lint}, with a representative process that gives 
the stated bound. 
\label{tbl:bestbounds}}
\end{center}
\end{minipage}
\end{center}
\end{table}

For interactions with many correlated couplings, an improved unitarity bound can be found by diagonalizing the transition matrix element.  For example, for the six point interaction 
\[
\frac{m_h^2}{16v^4}\de_3\, \vec{G}^6 = \frac{m_h^2}{16v^4}\de_3\, \left(G_1^2+G_2^2+G_3^2\right)^3,
\]
we expect the best scattering channel to appear for a specific custodial $SU(2)$ representation.  Focusing on the 3 to 3 processes, we use the basis 
\[
\left(\{3,0,0\}\; \{0,3,0\}\; \{0,0,3\}\; \{2,1,0\}\; \{2,0,1\}\; \{1,2,0\}\; \{0,2,1\}\; \{1,0,2\}\; \{0,1,2\}\;  \{1,1,1\}\right)^T
\]
where $\{n_1,n_2,n_3\}$ represents the number of goldstones $\{G_1,G_2,G_3\}$ in the state.  The transition matrix is 
\[
\left(\begin{array}{cccccccccc}
5&0&0&0&0&\sqrt{3}&0&\sqrt{3}&0&0\\
0&5&0&\sqrt{3}&0&0&0&0&\sqrt{3}&0\\
0&0&5&0&\sqrt{3}&0&\sqrt{3}&0&0&0\\
0&\sqrt{3}&0&3&0&0&0&0&1&0\\
0&0&\sqrt{3}&0&3&0&1&0&0&0\\
\sqrt{3}&0&0&0&0&3&0&1&0&0\\
0&0&\sqrt{3}&0&1&0&3&0&0&0\\
\sqrt{3}&0&0&0&0&1&0&3&0&0\\
0&\sqrt{3}&0&1&0&0&0&0&3&0\\
0&0&0&0&0&0&0&0&0&2
\end{array}\right) \frac{3m_h^2 \de_3}{512\pi^3 v^4} E^2
\]
which can be diagonalized to get a matrix with eigenvalues $7\cdot \frac{3m_h^2 \de_3}{512\pi^3 v^4} E^2
, 2\cdot \frac{3m_h^2 \de_3}{512\pi^3 v^4}E^2$ with multiplicity 3 and 7 respectively, which are the $I=1,3$ scattering channels.  Utilizing the larger eigenvalue for the $I=1$ channel leads to an optimized unitarity bound for 3 $G$ to 3 $G$ scattering of $13.4 \TeV/\sqrt{|\de_3|}$.  A similar analysis can also be performed for the $h\vec{G}^4$ interaction.  Analyzing the allowed 2 to 3 transition matrix, one finds eigenvalues of  $ \frac{15m_h^2 \de_3}{64\sqrt{2}\pi^2 v^3} E, \frac{3\sqrt{5}m_h^2 \de_3}{64\pi^2 v^3} E,$ and $\frac{3m_h^2 \de_3}{32\sqrt{2}\pi^2 v^3} E$ for the $I=0,1,2$ channels.  The best unitarity bound of $57.4 \TeV/|\de_3|$ comes from the $I=0$ channel.       
\bibliographystyle{utphys}
\bibliography{Higgs_Unitarity}

\end{document}